# Verbal Autopsy Methods with Multiple Causes of Death

**Gary King and Ying Lu**

*Abstract.* Verbal autopsy procedures are widely used for estimating cause-specific mortality in areas without medical death certification. Data on symptoms reported by caregivers along with the cause of death are collected from a medical facility, and the cause-of-death distribution is estimated in the population where only symptom data are available. Current approaches analyze only one cause at a time, involve assumptions judged difficult or impossible to satisfy, and require expensive, time-consuming, or unreliable physician reviews, expert algorithms, or parametric statistical models. By generalizing current approaches to analyze multiple causes, we show how most of the difficult assumptions underlying existing methods can be dropped. These generalizations also make physician review, expert algorithms and parametric statistical assumptions unnecessary. With theoretical results, and empirical analyses in data from China and Tanzania, we illustrate the accuracy of this approach. While no method of analyzing verbal autopsy data, including the more computationally intensive approach offered here, can give accurate estimates in all circumstances, the procedure offered is conceptually simpler, less expensive, more general, as or more replicable, and easier to use in practice than existing approaches. We also show how our focus on estimating aggregate proportions, which are the quantities of primary interest in verbal autopsy studies, may also greatly reduce the assumptions necessary for, and thus improve the performance of, many individual classifiers in this and other areas. As a companion to this paper, we also offer easy-to-use software that implements the methods discussed herein.

## 1. INTRODUCTION

National and international policymakers, public health officials, and medical personnel need information about the global distribution of deaths by cause in order to set research goals, budgetary priorities and ameliorative policies. Yet, only 23 of the world's 192 countries have high-quality death registration data, and 75 have no cause-specific mortality data at all (Mathers et al., 2005). Even if we include data of dubious quality, less than a third of the deaths that occur worldwide each year have a cause certified by medical personnel (Lopez et al., 2000).

*Gary King is David Florence Professor of Government, Department of Government, Harvard University, Institute for Quantitative Social Science, Cambridge, Massachusetts 02138, USA (e-mail: king@harvard.edu). Ying Lu is Assistant Professor, Departments of Sociology and Political Science, University of Colorado at Boulder, Boulder, Colorado 80309, USA (e-mail: Ying.Lu@colorado.edu).*

Verbal autopsy is a technique "growing in importance" (Sibai et al., 2001) for estimating the cause-of-death distribution in populations without vital registration or other medical death certification. It involves collecting information about symptoms (including signs and other indicators) from the caretakers of each of a randomly selected set of deceased in some population of interest, and inferring the cause of death. Inferences in these data are extrapolated either by physicians from their prior experiences or by statistical analysis of a second data set from a nearby hospital where information on symptoms from caretakers as well as validated causes of death are available.

Verbal autopsy studies are now widely used throughout the developing world to estimate cause-specific mortality, and are increasingly being used for disease surveillance and sample registration (Setel et al., 2005). Verbal autopsy is used on an ongoing basis and on a large scale in India and China, and in 36 demographic surveillance sites around the world (Soleman, Chandramohan and Shibuya, 2005). The technique has also proven useful in studying risk factors for specific diseases, infectious disease outbreaks, and the effects of public health interventions (Anker, 2003; Pacque-Margolis et al., 1990; Soleman, Chandramohan and Shibuya, 2006).

Until now, the most commonly used method has been physician review of symptoms with no additional validation sample. This approach can be expensive as it involves approximately three physicians, each taking 20–30 minutes to review symptoms and classify each death. To reduce the total time necessary, more physicians can be hired and work in parallel. Because judgments by these doctors are highly sensitive to their priors (when a Kansas doctor hears "fever and vomiting," malaria would not be her first thought), physicians need to come from local areas. This can pose difficult logistical problems because physicians in these areas are typically in very short supply, as well as serious ethical dilemmas since doctors are needed in the field for treating patients. Physician review also poses scientific problems since, although scholars have worked hard at increasing inter-physician reliability for individual studies, the cross-study reliability of this technique has remained low. Attempts to formalize physician reviews via expert-created deterministic algorithms are reliable by design, but appear to have lower levels of validity, in part because many diseases are not modeled explicitly and too many decisions need to be made.

Inferences from verbal autopsy data would thus seem ripe for adding to the growing list of areas where radically empirical approaches imbued with the power of modern statistics dominate human judgments by local experts (Dawes, Faust and Meehl, 1989). Unfortunately, the parametric statistical modeling that has been used in this area (known in the field as "data-derived techniques") has suffered from low levels of agreement with verified causes of death and is complicated for large numbers of causes. In practice, the choice of model has varied with almost every application. We attempt to rectify this situation.

In this article, we describe the current verbal autopsy approaches and the not always fully appreciated assumptions underlying them. We show that a key problem researchers have in satisfying most of the assumptions in real applications can be traced to the constraint existing methods impose by requiring the analysis of only one cause of death at a time. We generalize current methods to allow many causes of death to be analyzed simultaneously. This simple generalization turns out to have some considerable advantages for practice, such as making it unnecessary to conduct expensive physician reviews, specify parametric statistical models that predict the cause of death, or build elaborate expert algorithms. Although the missing (cause of death) information guarantees that verbal autopsy estimates always have an important element of uncertainty, the new approach offered here greatly reduces the unverified assumptions necessary to draw valid inferences. As a companion to this article, we are making available easy-to-use, free and open source software that implements all our procedures.

The structure of the inferential problem we study can also be found in application areas fairly distant from our verbal autopsy applications. Some version of the methods we discuss may be of use in these areas as well. For example, a goal of paleodemography is to estimate the age distribution in a large sample of skeletons from measurements of their physical features by using a small independent reference group where validated ages are available and skeletal features are also measured (Hoppa and Vaupel, 2002). Our methods seem to have already proven useful for estimating the proportion of text documents in each of a set of given categories, using a smaller reference set of text documents hand coded into the same categories (Hopkins and King, 2007) Also, as we show in Section 8, the methods introduced here imply that individual level classifiers can greatly reduce the assumptions necessary for accurate generalization to test sets with different distributional characteristics.

## 2. DATA DEFINITIONS AND INFERENTIAL GOALS

Denote the cause of death $j$ (for possible causes $j = 1, \ldots, J$) of individual $i$ as $D_i = j$. Bereaved relatives or caretakers are asked about each of a set of symptoms (possibly including signs or other indicators) experienced by the deceased before death. Each symptom $k$ (for possible symptoms $k = 1, \ldots, K$) is reported by bereaved relatives to have been present, which we denote for individual $i$ as $S_{ik} = 1$, or absent, $S_{ik} = 0$. We summarize the set of symptoms reported about an individual death, $\{S_{i1}, \ldots, S_{iK}\}$, as the vector $\mathbf{S}_i$. Thus, the cause of death $D_i$ is one variable with many possible values, whereas the symptoms $\mathbf{S}_i$ constitute a set of dichotomous variables.

Data come from two sources. The first is a hospital or other validation site, where both $\mathbf{S}_i$ and $D_i$ are available for each individual $i$ ($i = 1, \ldots, n$). The second is the community or some population about which we wish to make an inference, where we observe $\mathbf{S}_\ell$ (but not $D_\ell$) for each individual $\ell$ ($\ell = 1, \ldots, L$). Ideally, the second source of data constitutes a random sample from a large population of interest, but it could also represent any other relevant target group.

The quantity of interest for our analysis is $P(D)$, the distribution of cause-specific mortality in the population. Public health scholars are not normally interested in the cause of death $D_\ell$ of any particular individual in the population [although some current methods require estimates of these as intermediate values to compute $P(D)$]. They are sometimes also interested in the cause-of-death distribution for subgroups, such as age, sex, region or condition. We return to the implications of our approach for individual level classifiers in Section 8.

The difficulty of verbal autopsy analyses is that the population cause-of-death distribution is not necessarily the same in the hospital where $D$ is observed. In addition, researchers often do not sample from the hospital randomly, and instead oversample deaths due to causes that may be rare in the hospital. Thus, in general, the cause-of-death distribution in our two samples cannot be assumed to be the same: $P(D) \neq P^h(D)$.

Since symptoms are *consequences* of the cause of death, the data generation process has a clear ordering: Each disease or injury $D = j$ produces some symptom profiles (sometimes called "syndromes" or values of $\mathbf{S}$) with higher probability than others. We represent these conditional probability distributions as $P^h(\mathbf{S}|D)$ for data generated in the hospital and $P(\mathbf{S}|D)$ in the population. Thus, since the distribution of symptom profiles equals the distribution of symptoms given deaths weighted by the distribution of deaths, the symptom distribution will not normally be observed to be the same in the two samples: $P(\mathbf{S}) \neq P^h(\mathbf{S})$.

Whereas $P(D)$ is a multinomial distribution with $J$ outcomes, $P(\mathbf{S})$ may be thought of as either a multivariate distribution of $K$ binary variables or equivalently as a univariate multinomial distribution with $2^K$ possible outcomes, each of which is a possible symptom profile. We will usually use the $2^K$ representation.

## 3. CURRENT ESTIMATION APPROACHES

The most widely used current method for estimating cause-of-eath distributions in verbal autopsy data is physician review. What appears to be the best practice among the current statistical approaches used in the literature is the following multistage estimation strategy.

1. Choose a cause of death, which we here refer to as cause of death $D = 1$, apply the remaining steps to estimate $P(D = 1)$, and then repeat for each additional cause of interest (changing 1 to 2, then 3, etc.).

2. Using hospital data, develop a method of using a set of symptoms $\mathbf{S}$ to create a prediction for $D$, which we label $\hat{D}$ (and which takes on the value 1 or not 1). Some do this directly using informal, qualitative or deterministic prediction procedures, such as physician review or expert algorithms. Others use formal statistical prediction methods (called "data-derived algorithms" in the verbal autopsy literature), such as logistic regression or neural networks, which involve fitting $P^h(D|\mathbf{S})$ to the data and then turning it into a 0/1 prediction for an individual. Typically this means that if the estimate of $P^h(D = 1|\mathbf{S})$ is greater than 0.5, set the prediction as $\hat{D} = 1$ and otherwise set $\hat{D} \neq 1$. Of course, physicians and those who create expert algorithms implicitly calculate $P^h(D = 1 | \mathbf{S})$, even if they never do so formally.

3. Using data on the set of symptoms for each individual in the community, $\mathbf{S}_\ell$, and the same prediction method fit to hospital data, $P^h(D_\ell = 1 | \mathbf{S}_\ell)$, create a prediction $\hat{D}_\ell$ for all individuals sampled in the community ($\ell = 1, \ldots, L$) and average them to produce a preliminary or "crude" estimate of the prevalence of the disease of interest, $P(\hat{D} = 1) = \sum_{\ell=1}^{L} \mathbf{I}\hat{D}_{\ell=1}/L$.

4. Finally, estimate the *sensitivity*, $P^h(\hat{D} = 1 | D = 1)$, and *specificity*, $P^h(\hat{D} \neq 1 | D \neq 1)$, of the prediction method in hospital data via cross-validation and use it to "correct" the crude estimate and produce the final estimate:

$$
\begin{aligned}
(1) \quad & P(D = 1) \\
& = \frac{P(\hat{D} = 1) - [1 - P^h(\hat{D} \neq 1 | D \neq 1)]}{P^h(\hat{D} = 1 | D = 1) - [1 - P^h(\hat{D} \neq 1 | D \neq 1)]}.
\end{aligned}
$$

This correction, sometimes known as "back calculation," was first described in the verbal autopsy literature by Kalter (1992, Table 1) and originally developed for other purposes by Levy and Kass (1970). The correction is useful because the crude prediction, $P(\hat{D} = 1)$, can be inaccurate if sensitivity and specificity are not 100%.

A variety of creative modifications of this procedure have also been tried (Chandramohan, Maude, Rodrigues and Hayes, 1994). These include meta-analyses of collections of studies (Morris, Black and Tomaskovic, 2003), different methods of estimating $\hat{D}$, many applications with different sets of symptoms and different survey instruments (Soleman, Chandramohan and Shibuya, 2006), and other ways of combining the separate analyses from different diseases (Quigley et al., 2000; Boulle, Chandramohan and Weller, 2001). [See also work in statistics (Gelman, King and Liu, 1999) and political science (Franklin, 1989) that uses different approaches to methodologically related but substantively different problems.]

## 4. ASSUMPTIONS UNDERLYING CURRENT PRACTICE

The method described in Section 3 makes three key assumptions that we now describe. Then in the following section, we develop a generalized approach that reduces our reliance on the first assumption and renders the remaining two unnecessary.

The first assumption is that the sensitivity and specificity of $\hat{D}$ estimated from the hospital data are the same as those in the population:

$$
\begin{aligned}
(2) \quad & P(\hat{D} = 1 | D = 1) = P^h(\hat{D} = 1 | D = 1), \\
& P(\hat{D} \neq 1 | D \neq 1) = P^h(\hat{D} \neq 1 | D \neq 1).
\end{aligned}
$$

The literature contains much discussion of this assumption, the variability of estimates of sensitivity and specificity across sites, and good advice about controlling their variability (Kalter, 1992).

A less well-known but worrisome aspect of this first assumption arises from the choice of analyzing the $J$-category death variable as if it were a dichotomy. Because of the composite nature of the aggregated $D \neq 1$ category of death, we must assume that what makes up this composite is the same in the hospital and population. If it is not, then the required assumption about specificity (i.e., about the accuracy of estimation of this composite category) cannot hold in the hospital and population, even if sensitivity is the same. In fact, satisfying this assumption is more difficult than may be generally understood. To make this point, we begin with the decomposition of specificity, offered by Chandramohan, Setel and Quigley (2001) (see also Maude and Ross (1997)), as one minus the sum of the probability of different misclassifications times their respective prevalences:

$$
\begin{aligned}
(3) \quad & P(\hat{D} \neq 1 | D \neq 1) \\
& = 1 - \sum_{j=2}^{J} P(\hat{D} = 1 | D = j) \frac{P(D = j)}{P(D \neq 1)},
\end{aligned}
$$

which emphasizes the composite nature of the $D \neq 1$ category. Then we ask: *under what conditions can specificity in the hospital equal that in the population if the distribution of cause of death differs?* The mathematical condition can be easily derived by substituting (3) into each side of the second equation of (2) (and simplifying by dropping the "$1-$" on both sides):

$$
\begin{aligned}
(4) \quad & \sum_{j=2}^{J} P(\hat{D} = 1 | D = j) \frac{P(D = j)}{P(D \neq j)} \\
& = \sum_{j=2}^{J} P^h(\hat{D} = 1 | D = j) \frac{P^h(D = j)}{P^h(D \neq j)}.
\end{aligned}
$$

If this equation holds, then this first assumption holds. And if $J = 2$, this equation reduces to the first line of (2) and so, in that situation, the assumption is unproblematic.

However, for more than two diseases specificity involves a composite cause of death category. We know that the distribution of causes-of-death [the last factor on each side of (4)] differs in the hospital and population by design, and so the equation can hold only if a miraculous mathematical coincidence holds, whereby the probability of misclassifying each cause of death as the first cause occurs in a pattern that happens to cancel out differences in the prevalence of causes between the two samples. For example, this would not occur according to any theory or observation of mortality patterns offered in the literature. Verbal autopsy scholars recognize that some values of sensitivity and specificity are impossible when (1) produces estimates of $P(D = 1)$ greater than 1. They then use information to question the values of, or modify, estimates of sensitivity and specificity, but the problem is not necessarily due to incorrect estimates of these quantities and could merely be due to the fact that the procedure requires assumptions that are impossible to meet. In fact, *as the number of causes of death increases, the required assumption can only hold if sensitivity and specificity are each* 100%, which we know does not describe real data. [The text describes how this first assumption can be met by discussing specificity only with respect to cause of death 1. In the general case, (4) for all causes requires satisfying $\sum_j P(\hat{D} \neq j | D \neq j) - (J - 2) = \sum_j [P(\hat{D} \neq j | D \neq j) + P(\hat{D} = j | D = j)] P(D = j)$. For small $J > 2$, this will hold only if a highly unlikely mathematical coincidence occurs; for large $J$, this condition is not met in general unless sensitivity and specificity are 1 for all $j$.]

The second assumption is that the (explicit or implicit) model underlying the prediction method used in the hospital must also hold in the population: $P(D|\mathbf{S}) = P^h(D|\mathbf{S})$. For example, if logistic regression is the prediction method, we make this assumption by taking the coefficients estimated in hospital data and using them to multiply by symptoms collected in the population to predict the cause of death in the population. This is an important assumption, but not a natural one since the data generation process is the reverse: $P(\mathbf{S}|D)$. And most importantly, even if the identical data generation process held in the population and hospital, $P(\mathbf{S}|D) = P^h(\mathbf{S}|D)$, we would still have no reason to believe that $P(D|\mathbf{S}) = P^h(D|\mathbf{S})$ holds. The assumption might hold by luck, but coming up with a good reason why we should believe it holds in any real case seems unlikely.

This problem is easy to see by generating data from a regression model with $D$ as the explanatory variable and $\mathbf{S}$ as the simple dependent variable, and then regressing $\mathbf{S}$ on $D$: Unless the regression fits perfectly, the coefficients from the first regression do not determine those in the second. Similarly, when Spring comes, we are much more likely to see many green leaves; but visiting the vegetable section of the supermarket in the middle of the winter seems unlikely to cause the earth's axis to tilt toward the sun. Of course, it just *may* be that we can find a prediction method for which $P(D|\mathbf{S}) = P^h(D|\mathbf{S})$ holds, but knowing whether it does or even having a theory about it seems unlikely. It is also *possible*, with a small number of causes of death, that the sensitivity and specificity for the wrong model fit to hospital data could by chance be correct when applied to the population, but it is hard to conceive of a situation when we would know this ex ante. This is especially true given the issues with the first assumption: the fact that the composite $D \neq 1$ category is by definition different in the population and hospital implies that different symptoms will be required predictors for the two models, hence invalidating this assumption.

A final problem with the current approach is that the multistage procedure estimates $P(D = j)$ for each $j$ separately, but for the ultimate results to make any sense the probability of a death occurring due to some cause must be 100%: $\sum_{j=1}^J P(D = j) = 1$. This can happen if the standard estimation method is used, but it will hold only by chance.

## 5. AN ALTERNATIVE APPROACH

The key problem underlying the veracity of each of the assumptions in Section 4 can be traced to the practice of sequentially dichotomizing the $J$-category cause-of-death variable. In analyzing the first assumption, we learn that specificity cannot be equal in hospital and population data as the number of causes that make up the composite residual category gets large. In the second assumption, the practice of collapsing the relationship between $\mathbf{S}$ and $D$ into a dichotomous prediction, $\hat{D}$, requires making assumptions opposite to the data generation process and either a sophisticated statistical model, or an expensive physician review or set of expert algorithms, to summarize $P(D|\mathbf{S})$. And finally, the estimated cause-of-death probabilities do not necessarily sum to 1 in the existing approach precisely because $D$ is dichotomized in multiple ways and each dichotomy is analyzed separately.

Dichotomization has been used in each case to simplify the problem. However, we show in this section that most aspects of the assumptions with the existing approach are unnecessary once we treat the $J$-category cause-of-death variable as having $J$ categories. Moreover, it is simpler conceptually than the current approach. We begin by *reformulating* the current approach so it is more amenable to further analysis and then *generalizing* it to the $J$-category case.

*Reformulation.* Under the current method's assumption that sensitivity and specificity are the same in the hospital and population, we can rearrange the back-calculation formula in (1) as

$$
\begin{aligned}
(5) \quad P(\hat{D} = 1) = {}& P(\hat{D} = 1 | D = 1) P(D = 1) \\
& + P(\hat{D} = 1 | D \neq 1) P(D \neq 1),
\end{aligned}
$$

and rewrite (5) in equivalent matrix terms as

$$
(6) \qquad \underset{2 \times 1}{P(\hat{D})} = \underset{2 \times 2}{P(\hat{D}|D)} \underset{2 \times 1}{P(D)}
$$

where the extra notation indicates the dimension of the matrix or vector. So $P(\hat{D})$ and $P(D)$ are now both $2 \times 1$ vectors, and have elements $[P(\hat{D} = 1), P(\hat{D} \neq 1)]'$ and $[P(D = 1), P(D \neq 1)]'$, respectively; and $P(\hat{D}|D)$ is a $2 \times 2$ matrix where

$$
\underset{2 \times 2}{P(\hat{D}|D)} = \begin{pmatrix} P(\hat{D} = 1 | D = 1) & P(\hat{D} = 1 | D \neq 1) \\ P(\hat{D} \neq 1 | D = 1) & P(\hat{D} \neq 1 | D \neq 1) \end{pmatrix}.
$$

Whereas (1) is solved for $P(D = 1)$ by plugging in values for each term on the right-hand side, (6) is solved for $P(D)$ by linear algebra. Fortunately, the linear algebra required is simple and well known from the least squares solution in linear regression. We thus recognize $P(\hat{D})$ as taking the role of a "dependent variable," $P(\hat{D}|D)$ as two "explanatory variables," and $P(D)$ as the coefficient vector to be solved for. Applying least squares yields an estimate of $P(D)$, the first element of which, $P(D = 1)$, is exactly the same as that in (1). Thus far, only the mathematical representation has changed; the assumptions, intuitions and estimator remain identical to the existing method described in Section 3.

*Generalization.* The advantage of switching to matrix representations is that they can be readily generalized, which we do now in two important ways. First, we drop the modeling necessary to produce the cause-of-death for each individual $\hat{D}$, and use $\mathbf{S}$ in its place directly. And second, we do not dichotomize $D$ and instead treat it as a full $J$-category variable. We implement both generalizations via a matrix expression that is the direct analogue of (6):

$$
(7) \qquad \underset{2^K \times 1}{P(\mathbf{S})} = \underset{2^K \times J}{P(\mathbf{S}|D)} \underset{J \times 1}{P(D)}.
$$

The quantity of interest in this expression remains $P(D)$. Although we use the better nonparametric estimation methods (described in the Appendix), we could in principle estimate $P(\mathbf{S})$ by direct tabulation, by simply counting the fraction of people in the population who have each symptom profile. Since we do not observe and cannot directly estimate $P(\mathbf{S}|D)$ in the community (because $D$ is unobserved), we estimate it from the hospital data via (nonparametric) tabulation and assume $P(\mathbf{S}|D) = P^h(\mathbf{S}|D)$. We estimate $P^h(\mathbf{S}|D = j)$ for each cause-of-death $j$ the same way as we do for $P(\mathbf{S})$.

The only assumption required for connecting the two samples is $P(\mathbf{S}|D) = P^h(\mathbf{S}|D)$, which is natural as it directly corresponds to the data generation process. We do not assume that $P(\mathbf{S})$ and $P^h(\mathbf{S})$ are equal, $P(D)$ and $P^h(D)$ are equal, or $P(D|\mathbf{S})$ and $P^h(D|\mathbf{S})$ are equal. In fact, prediction methods for estimating $P(D|\mathbf{S})$ or $\hat{D}$ are entirely unnecessary here, and so unlike the current approach, we do not require parametric statistical modeling, physician review or expert algorithms.

We solve (7) for $P(D)$ directly. This can be done conceptually using least squares. That is, $P(\mathbf{S})$ takes the role of a "dependent variable," $P(\mathbf{S}|D)$ takes the role of a matrix of $J$ "explanatory variables," each column corresponding to a different cause-of-death, and $P(D)$ is the "coefficient vector" with $J$ elements for which we wish to solve. We also modify this procedure to ensure that the estimates of $P(D)$ are each between zero and 1 and together sum to 1 by changing least squares to constrained least squares (see the Appendix).

Although producing estimates from this expression involves some computational complexities, which we describe in Section 7 and the Appendix, this is a single equation procedure that is conceptually far simpler than current practice. As described in Section 3, the existing approach requires four steps, applied sequentially to each cause-of-death. In contrast, estimates from our proposed alternative only require understanding each term in (7) and solving for $P(D)$.

## 6. ILLUSTRATIONS IN DATA FROM CHINA AND TANZANIA

Since deaths are not observed in populations for which verbal autopsy methods are used, realistic validation of any method is, by definition, difficult or impossible (Gajalakshmi and Peto, 2004). We attempt to validate our method in two separate ways in data from China and Tanzania.

*China.* We begin with an analysis of 2822 registered deaths from hospitals in urban China collected and analyzed by Alan Lopez and colleagues (see, most recently, Yang et al., 2005). Thirteen causes of death were coded, and 56 (yes or no) symptoms were elicited from caretakers. We conducted three separate analyses with these data. We designed the first test to meet the assumptions of our method by randomly splitting these data into halves. Although all these data were collected in hospitals, where we observe both **S** and $D$, we label the first random set "hospital data," for which we use both **S** and $D$, and the second "population data," for which we *only* use **S** during estimation. We emulate an actual verbal autopsy analysis by using these data to estimate the death frequency distribution, $P(D)$, in the "population data." Finally, for validation, we unveil the actual cause-of-death variable for the "population data" that were set aside during the analysis and compare it to our estimates.

The estimates appear in the top panel of the left graph of Figure 1, which plots on the horizontal axis a direct sample estimate—the proportion of the sample from the population dying from each of 13 causes—and on the vertical axis an estimate from our verbal autopsy method. (This direct estimator is not normally feasible in verbal autopsy studies because of the impossibility of obtaining medically verified cause-of-death data in the community.) Since both are sample-based estimates, and thus both are measured with error, if our method predicted perfectly, all points would fall approximately on the 45° line. Clearly, the fit of our estimates to the direct estimates of the truth is fairly close, with no clear pattern in deviations from the line. The bottom panel of this graph portrays the difference between our estimates and the direct sample estimates, along with a 95% confidence interval for the difference. Almost all confidence intervals of the errors cover no difference (portrayed as a horizontal line), which indicates approximately accurate coverage.

For a more stringent test of our approach, we split the same sample into 1409 observations from hospitals in three cities (Beijing, Chengdu and Wuhan) and 1413 observations from hospitals in another three cities (Haierbin, Guangzhou and Shanghai). We then let each group take a turn playing the role of the "population" sample (with known cause-of-death that we use only for validation) and the other as the actual hospital sample. These are more difficult tests of our method than would be necessary in practice, since researchers would normally collect hospital data from a facility physically closer to, part of, and more similar to the population to which they wish to infer.

The right two graphs in Figure 1 give results from this test in the same format as for the random split on the left. The middle graph estimates the cause-of-death distribution of our first group of sample cities from the second group, whereas the right graph does the reverse. The fit between the directly estimated true death proportions and our estimates in both is slightly worse than for the left graph, where our assumptions were true by construction, but predictions in both are still excellent. Again, almost all of the 95% confidence intervals for the difference between our estimator and the direct sample estimate cross the zero line (see the bottom of each graph).

*Tanzania.* We also analyze cause-specific adult mortality from a verbal autopsy study in Tanzania (see Setel et al., 2006). The data include 1261 hospital deaths and 282 deaths from the general population, about which 51 symptoms questions and 13 causes of death were collected. The unusual feature of these data is that all the population deaths have medically certified causes, and so we can set aside that information and use it to validate our approach. We again use **S** and $D$ from the hospital and **S** from the population and attempt to estimate $P(D)$ in the population, using $D$ from the population only for validation after the estimation is complete.

The results appear in Figure 2 in the same format as the China data. We constructed randomly split data on the left and an actual prediction to the community for the graph on the right. The results are similar to those in China, where the point estimates appear roughly spread around the 45° line, indicating, in this very different context, that the fit is approximately as good—and again better for the random split than the actual forecast. The confidence intervals of the differences between the direct estimate and our estimate, in the bottom panel, are larger than for the China data due to the smaller target population used to estimate $P(\mathbf{S})$, but almost all the intervals cross zero.

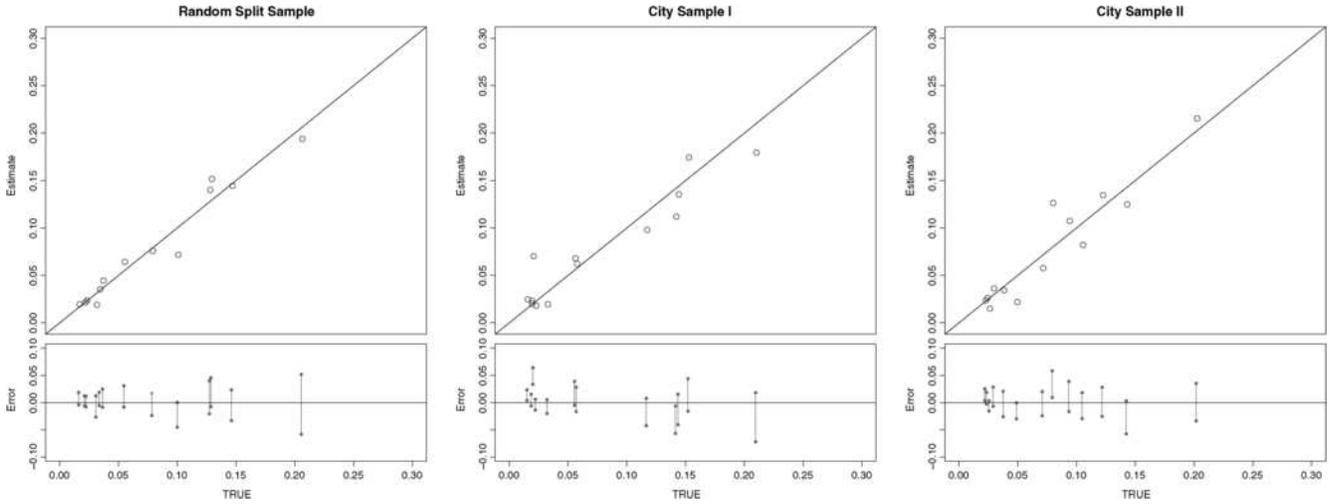

FIG. 1. *Validation in China. For validation, we break data with known causes of death into randomly split halves (left graph) and arbitrarily by groups of hospitals in different cities (the two right graphs). The top panel in each graph plots a direct estimate of cause-specific mortality horizontally by the estimate from our method vertically. The bottom panel of each graph contains 95% confidence intervals of the difference between our estimator and the direct estimate, both of which are measured with error; almost all of these vertical lines cross the zero difference point marked by a horizontal line.*

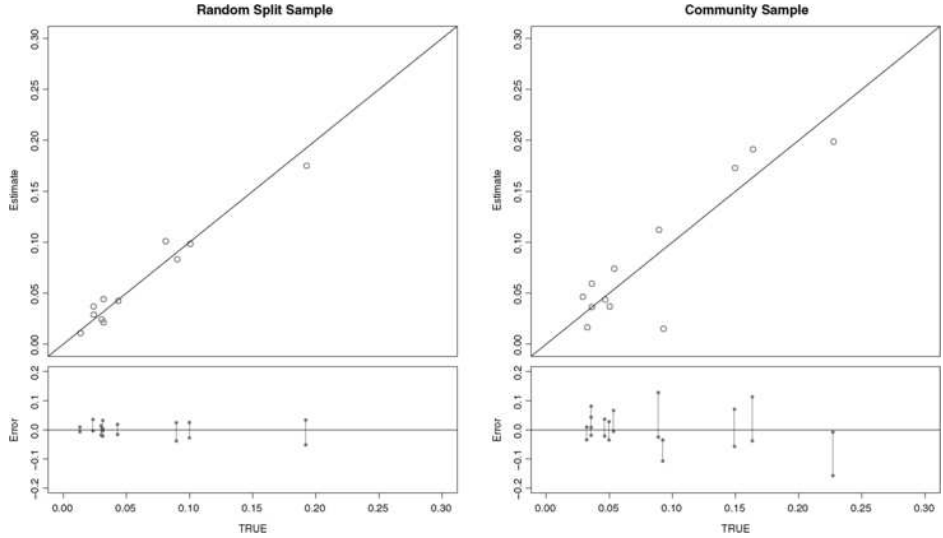

FIG. 2. *Validation in Tanzania. Each graph plots the (normally unknown) direct estimate of cause-specific mortality horizontally and estimates from our method vertically. This is done for data based on a random split, where our assumptions are true by construction, on the left and for predictions of the community sample based on hospital sample on the right. The bottom panel of each graph gives the 95% confidence interval of the difference between the direct estimate and our estimate, both of which are measured with error.*

The variance of the direct sampling estimator, $\bar{D}_j$, is approximately $\bar{D}_j(1 - \bar{D}_j)/n$, and thus varies with category size. Uncertainty estimates from our approach are computed by bootstrapping, and of course also vary by category size. The 95% confidence interval from our estimator is on average across categories 50% wider than the direct sampling estimator in the China data and 25% wider in the Tanzania data. Obviously, the reason verbal autopsy procedures are necessary is that direct sampling estimates of the cause-of-death in the population are unobtainable, and so these numbers summarize the necessary costs incurred for this lack of information. Of course, compared to the huge costs of complete national vital registration systems, this is a trivial difference.

## 7. INTERPRETATION

We offer five interpretations of our approach. First, since $\mathbf{S}$ contains $K$ dichotomous variables and thus $2^K$ symptom profiles, $P(\mathbf{S})$ and $P(\mathbf{S}|D)$ have $2^K$ rows, which take the role of "observations" in the linear expression in (7). By analogy to linear regression, where more observations make for more efficient estimates (i.e., with lower variances), we can see clearly here that having additional symptoms that meet the assumptions of verbal autopsy studies will decrease the variance, but not affect the bias, of our estimates of cause-specific mortality.

Second, when the number of symptoms is large, direct tabulation can produce an extremely sparse matrix for $P(\mathbf{S})$ and $P(\mathbf{S}|D)$. For example, our data from China introduced in Section 6 have 56 symptoms, and so we would need to sort the $n = 1411$ observations collected from the population into $2^{56}$ categories, which number more than 72 quadrillion. Reliable estimation by direct tabulation in this case is obviously infeasible. In practice, we only need to keep the symptom profiles that actually appear in both the hospital and population data sets, but even this can be sparsely populated and so will leave few nonzero rows available in both. We thus develop an easy computational solution to this problem in the Appendix based on a variant of discrete kernel smoothing, which involves using random subsets of symptoms, solving (7) for each, and averaging. The difference here is that unlike the usual applications of kernel smoothing, which reduce variance at the expense of some bias, our procedure would appear to reduce both bias and variance here.

Third, the key statistical assumption of the method connecting the two samples is that $P(\mathbf{S}|D) = P^h(\mathbf{S}|D)$. If this expression holds in sample, then our method (and indeed every subset calculation) will yield the true $P(D)$ population proportions exactly, regardless of the degree of sparseness. If the assumption instead holds only in the population from which the observed data are drawn, then our approach will yield statistically consistent estimates of the population density $P(D)$. If, in addition, subset sizes are small enough, then we find through simulation that our estimates are approximately unbiased.

Substantively, this key assumption would fail, for example, for symptoms that doctors make relatives more aware of in the hospital; following standard advice for writing survey questions simply and concretely can eliminate many of these issues. Another way this assumption can be violated would be if hospitals keep patients alive for certain diseases longer than they would be kept alive in the community, and as a result they experience different symptoms. In these examples, and others, an advantage of our approach, compared to approaches which model $P(D|\mathbf{S})$, is that researchers have the freedom to drop symptoms that would seem to severely violate the assumption.

Fourth, a reasonable question is whether expert knowledge from physicians or others could somehow be used to improve our estimation technique. This is indeed possible, via a Bayesian extension of our approach that we have also implemented. However, in experimenting with our methods with verbal autopsy researchers, we found few sufficiently confident of the information available to them from physicians and others that they would be willing to add Bayesian priors to the method described here. We thus do not develop our full Bayesian method here, but we note that if accurate prior information does exist in some application and were used, it would improve our estimates (see also Sibai et al. 2001).

Finally, the new approach represents a major change in perspective in the verbal autopsy field. The essential goal of the existing approach is to marshal the best methods to use $\mathbf{S}$ to predict $D$. The thought is that if we can only nail down the "correct" symptoms, and use them to generate predictions with high sensitivity and specificity, we can get the right answer. There are corrections for when this fails, of course, but the conceptual perspective involves developing a *proxy* for $D$. That proxy can be well-chosen symptoms or symptom profiles, or a particular aggregation of profiles as $\hat{D}$. The existing literature does not seem to offer methods for highly accurate predictions of $D$, even before we account for the difficulties in ascertaining the success of classifiers (Hand, 2006). Our alternative approach would also work well if symptoms or symptom profiles are chosen well enough to provide accurate predictions of $D$, but accurate predictions are unnecessary. In fact, choosing symptoms with higher sensitivity and specificity would not reduce bias in our approach, but in the existing approach they are required for unbiasedness except for lucky mathematical coincidences.

Instead of serving as proxies, symptoms in the new approach are only meant to be observable *implications* of $D$, and any subset of implications is fine. They need not be biological assays or in some way fundamental to the definition of the disease or injury or an exhaustive list. Symptoms need to occur with particular patterns more for some causes of death than others, but bigger differences do not help reduce bias (although they may reduce the variance). The key assumption of our approach is $P(\mathbf{S}|D) = P^h(\mathbf{S}|D)$. Since $\mathbf{S}$ is entirely separable into individual binary variables, we are at liberty to choose symptoms in order to make this assumption more likely to hold. The only other criteria for choosing symptoms, then, are the usual rules for reducing measurement error in surveys, such as reliability, question-ordering effects, question wording, and ensuring that different types of respondents interpret the same symptom questions in similar ways. Other previously used criteria, such as sensitivity, specificity, false positive or negative rates, or other measures of predictability, are not of as much relevance as criteria for choosing symptom questions.

## 8. IMPLICATIONS FOR INDIVIDUAL CLASSIFIERS

We now briefly discuss the implications of our work for classification of the cause of each individual death. As the same results would seem to have broader implications for the general problem of individual classification in a variety of applications, we generalize the discussion here but retain our notation with $\mathbf{S}$ referring to what is called in the classifier literature features or covariates and $D$ denoting category labels.

As Hand (2006, page 7) emphasizes, "Intrinsic to the classical supervised classification paradigm is the assumption that the data in the design set are randomly drawn from the same distribution as the points to be classified in the future." In other words, individual classifiers make the assumption that the *joint* distribution of the data is the same in the unlabeled (community) set as in the labeled (hospital) set, $P(\mathbf{S}, D) = P^h(\mathbf{S}, D)$, a highly restrictive and often unrealistic condition. If $P(D|\mathbf{S})$ fits exceptionally well (i.e., with near 100% sensitivity and specificity), then this common joint distribution assumption is not necessary, but classifiers rarely fit that well.

In verbal autopsy applications, assuming common joint distributions or nearly perfect predictors is almost always wrong. Hand (2006) gives many reasons why these assumptions are wrong as well in many other types of classification problems. We add to his list a revealing fact suggested by our results above: Because $P(\mathbf{S})$ and $P^h(\mathbf{S})$ are directly estimable from the unlabeled and labeled sets, respectively, these features of the joint distribution can be directly compared and this one aspect of the common joint distribution assumption can be tested directly. Of course, the fact that this assumption can be tested also implies that this aspect of the common joint distribution assumption need not be made in the first place. In particular, we have shown above that we need not assume that $P(\mathbf{S}) = P^h(\mathbf{S})$ or $P(D) = P^h(D)$ when trying to estimate the aggregate proportions. We show here that these assumptions are also unnecessary in individual classifications.

Thus, instead of assuming a common joint distribution between the labeled and unlabeled sets, we make the considerably less restrictive assumption that only the *conditional* distributions are the same: $P(\mathbf{S}|D) = P^h(\mathbf{S}|D)$. (As above, we get the needed joint distribution in the unlabeled set by multiplying this conditional distribution estimated from the labeled set by the marginal distribution $P(\mathbf{S})$ estimated directly from the unlabeled set.) Thus, to generalize our results to apply to individual classification, which requires an estimate of $P(D_\ell = j|\mathbf{S}_\ell = \mathbf{s}_\ell)$, we use Bayes theorem:

$$(8) \quad \begin{aligned} & P(D_\ell = j|\mathbf{S}_\ell = \mathbf{s}_\ell) \\ & = \frac{P(\mathbf{S}_\ell = \mathbf{s}_\ell|D_\ell = j)P(D_\ell = j)}{P(\mathbf{S}_\ell = \mathbf{s}_\ell)}. \end{aligned}$$

We propose to use this by taking $P(\mathbf{S}_\ell = \mathbf{s}_\ell|D_\ell = j)$ from the labeled set, the estimated value of $P(D_\ell = j)$ from the procedure described in Section 5, and $P(\mathbf{S}_\ell = \mathbf{s}_\ell)$ directly estimated nonparametrically from the unlabeled set, also as in Section 5. As with our procedure, we use subsets of $\mathbf{S}$ and average different estimates of $P(D_\ell|\mathbf{S}_i = \mathbf{s}_i)$, although this time the averaging is via committee methods since each

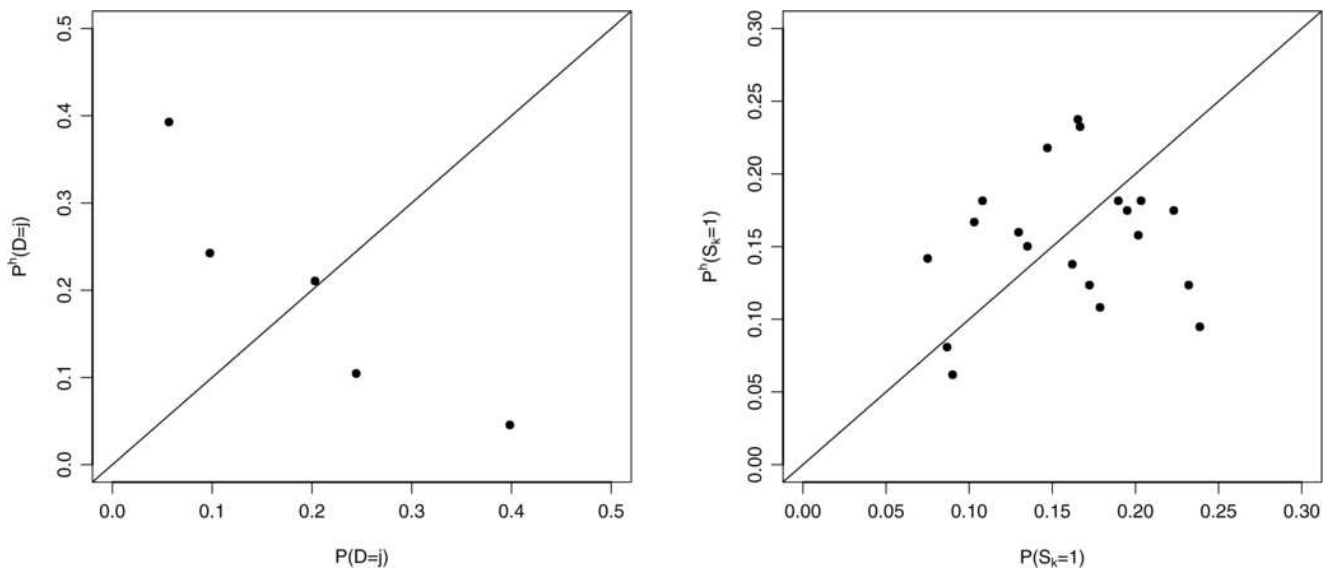

Fɪɢ. 3. *Simulated data. For both the proportion of observations in each category (in the left panel) and the proportion with each feature present (in the right panel), the labeled set is very different from the unlabeled target population of interest. These data would violate the assumptions underlying most existing classifiers.*

subset implies a different model [with the result constrained so that the individual classifications aggregate to the $\hat{P}(D)$ estimate]. Each of these lower-dimensional subsets (labeled "sub") also imply easier-to-satisfy assumptions than the full conditional relationship, $P(\mathbf{S}_{\text{sub}}|D) = P^h(\mathbf{S}_{\text{sub}}|D)$. The key advantage of this approach is that it uses more information from the unlabeled set—that is, $P(\mathbf{S})$—than existing classifiers. If the unlabeled set was generated from the population such that the distribution of the values of the features is informative, then this alternative approach can greatly improve estimation accuracy.

We illustrate the power of these results with a simple simulation. For simplicity, we assume that features are independent conditional on the category labels in the labeled set, $P^h(\mathbf{S} = \mathbf{s}|D) = \prod_{k=1}^{K} P(S_k = s_k|D)$, which is empirically reasonable except for heterogeneous residual categories. We then simulate data, with 5 (disease) categories, 20 (symptom) features, and 3000 observations in the labeled (hospital) and unlabeled (community) sets. We generate the data so they have very different marginal distributions for $P(\mathbf{S})$ and $P(D)$. Figure 3 gives these marginal distributions, plotting the unlabeled set values horizontally and labeled set vertically; note that few points are near the 45° line. These data are generated to violate the common joint distribution assumptions of all existing standard classifiers, but still meet the less restrictive conditional distribution assumption.

We then run a standard support vector machine classifier (Chang and Lin, 2001) on the simulated data, which classifies only 40.5% of the observations correctly. In contrast, our simple nonparametric alternative classifies 59.8% of the same observations correctly. The key advantage here is coming from the adjustment of the marginals to fit $\hat{P}(D)$ in the "unlabeled" set. We can see this by viewing the aggregate results. These appear in Figure 4, with the truth plotted horizontally and estimates vertically. Note that our estimates (plotted with black disks) are much closer to the 45° line for every true value than the SVM estimates (plotted with open circles).

This section illustrates only the general implications of our strategy for individual classification. It should be straightforward to extend these results to provide a simple but powerful correction to any existing classifier, as well as a more complete nonparametric classifier.

## 9. CONCLUDING REMARKS

By reducing the assumptions necessary for valid inference and making it possible to model all diseases simultaneously, the methods introduced here make it possible to extract considerably more information from verbal autopsy data, and as a result can produce more accurate estimates of cause-specific mortality rates. Since our approach makes physician reviews, expert algorithms and parametric statistical models unnecessary, it costs considerably less to implement and is easier to replicate in different settings and by different researchers. The resulting increased accuracy of our relatively automated statistical approach, compared to existing methods which require many more ad hoc human judgments, is consistent with a wide array of research in other fields (Dawes, Faust and Meehl, 1989).

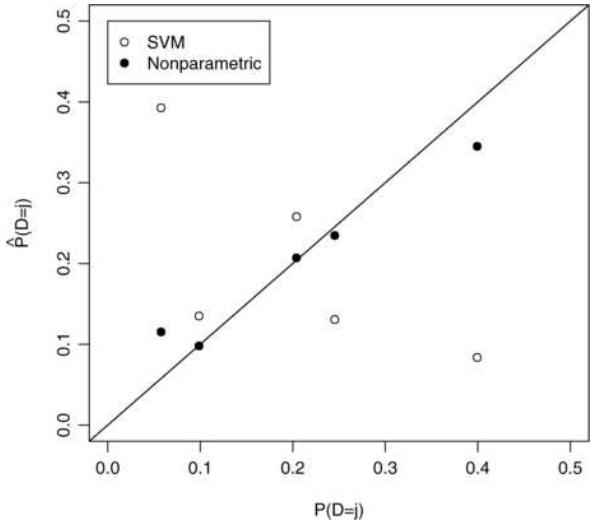

Fig. 4. *Individual-level classification by support vector machine (open circles) and our improved nonparametric alternative (closed disks). Despite the differences between the labeled and unlabeled sets in Figure 3, our approach generates better aggregate results than the standard support vector machine classifier.*

Even with the approach offered here, many issues remain. For example, to estimate the distribution of death by age, sex or condition with our methods requires separate samples for each group. To save money and time, the methods developed here could also be extended to allow covariates, which would enable these group-specific effects to be estimated simultaneously from the same sample. A Bayesian approach could also be applied to borrow strength across these areas. A formal approach to choosing the smoothing parameter (the number of symptoms per subset) would be useful as well. In addition, scholars still need to work on reducing errors in eliciting symptom data from caregivers and validating the cause-of-death. Progress is needed on procedures for classifying causes of death and statistical procedures to correct for the remaining misclassifications, and on question wording, recall bias, question-ordering effects, respondent selection, and interviewer training for symptom data. Crucial issues also remain in choosing a source of validation data for each study similar enough to the target population so that the necessary assumptions hold, and in developing procedures that can more effectively extrapolate assumptions from hospital to population via appropriate hospital subpopulations, data collection from community hospitals, or medical records for a sample of deaths in the target population.

## APPENDIX: ESTIMATION METHODS

We now describe the details of our estimation strategy. Instead of trying to use all $2^K$ symptoms simultaneously, which will typically be infeasible given commonly used sample sizes, we recognize that only full rank subsets larger than $J$ with sufficient data are required. We thus sample many subsets of symptoms, estimate $P(D)$ in each, and average the results (or if prior information is available we use a weighted average). To choose subsets, we could draw directly from the $2^K$ symptom profiles, but instead use the convenient approach of randomly drawing $B$ ($B < K$) symptoms, which we index as $I(B)$, and use the resulting symptom subprofile. This procedure is mathematically equivalent to imposing a version of kernel smoothing on an otherwise highly sparse estimation task. (More advanced versions of kernel smoothing might improve these estimates further.)

We estimate $P(\mathbf{S}_{I(B)})$ using the population data, and $P(\mathbf{S}_{I(B)}|D)$ using the hospital data. Denote $Y = P(\mathbf{S}_{I(B)})$ and $X = P(\mathbf{S}_{I(B)}|D)$, where $Y$ is of length $n$, $X$ is $n \times J$, and $n$ is the subset of the $2^B$ symptom profiles that we observe. We obtain $P(D) \equiv \hat{\beta}$ by regressing $Y$ on $X$ under the constraint that elements of $\hat{\beta}$ fall on the simplex. The subset size $B$ should be chosen to be large enough to reduce estimation variance (and so that the number of observed symptom profiles among the $2^B$ possible profiles is larger than $J$) and small enough to avoid the bias that would be incurred from sparse counts used to estimate elements of $P(\mathbf{S}_{I(B)}|D)$. We handle missing data by deleting incomplete observations within each subset (another possibility would be model-based imputation). Although cross-validation can generate optimal choices for $B$, we find estimates of $P(D)$ to be relatively robust to choices of $B$ within a reasonable range. [When choosing $B$ via cross-validation from the hospital data, we use random subsets to separate this decision from the assumption that $P(\mathbf{S}|D) = P^h(\mathbf{S}|D)$.] We have experimented with nonlinear optimization procedures to estimate $P(D)$ directly, but it tends to be sensitive to starting values when $J$ is large. As an alternative, we developed the following estimation procedure, which tends to be much faster, more reliable, and accurate in practice.

We repeat the following two steps for each different subset of symptoms and then average the results. The two steps involve reparameterization, to ensure $\sum \beta_j = 1$, and stepwise deletion, to ensure $\beta_j > 0$.

1. To reparameterize, we follow this algorithm:
   (a) To impose a fixed value for some cause-of-death, $\sum \beta_j = c$, rewrite the constraint as $C\beta = 1$, where $C$ is a $J$-row vector of $\frac{1}{c}$. When none of the elements of $\beta$ are known a priori, $c = 1$. When we know some elements $\beta_i$, such as from another data source, the constraint on the rest of $\beta$ changes to $\sum_{j \neq i} \beta_j = c = 1 - \beta_i$.
   (b) Construct a $J - 1 \times J$ matrix $A$ of rank $J - 1$ whose rows are mutually orthogonal and also orthogonal to $C$, and so $CA^\top = 0$ and $AA^\top = I_{J-1}$. A Gram–Schmidt orthogonalization gives us a row-orthogonal matrix $G$ whose first row is $C$, and the rest is $A$.
   (c) Rewrite the regressor as $X = ZA + WC$, where $Z$ is $n \times J - 1$, $W$ is $n \times 1$ and $(W, Z)G = X$. Under the constraint $C\beta = 1$, we have $Y = X\beta = ZA\beta + WC\beta = Z\gamma + W$, where $\gamma = A\beta$, and $\gamma$ is a $J - 1$ vector.
   (d) Obtain the least square estimate $\hat{\gamma} = (Z^\top Z)^{-1} Z^\top (Y - W)$.
   (e) The equality-constrained $\beta$ is then $\hat{\beta} = G^{-1} \gamma^*$, where $G = (C, A)$, a $J \times J$ row-orthogonal matrix derived above, and $\gamma^* = (1, \hat{\gamma})$. This ensures that $C\hat{\beta} = 1$. Moreover, $\text{Cov}(\hat{\beta}) = G^{-1} \text{Cov}(\gamma^*)(G^\top)^{-1}$ (Thisted, 1988).
2. Then for stepwise deletion:
   (a) To impose nonnegativity, find the $\hat{\beta}_j < 0$ whose associated $t$-value is the biggest in absolute value among all $\hat{\beta} < 0$.
   (b) Remove the $j$th column of the regressor $X$, and go to the *reparameterization* step again to obtain $\hat{\beta}$ with the $j$th element coerced to zero.

Alternatively, we can view the estimation of $\beta$ to be a constrained optimization problem and use the dual method to solve the strictly convex quadratic programs. Finally, our estimate of $P(D)$ can be obtained by averaging over the estimates based on each subset of symptoms. The associated standard error can be estimated by bootstrapping over the entire algorithm. Subsetting is required because of the size of the problem, but because $\mathbf{S}$ can be subdivided and our existing assumption $P(\mathbf{S}|D) = P^h(\mathbf{S}|D)$ implies $P(\mathbf{S}_{I(B)}|D) = P^h(\mathbf{S}_{I(B)}|D)$ in each subset, no bias is introduced. In addition, although the procedure is statistically consistent (i.e., as $n \to \infty$ with $K$ fixed), the procedure is approximately unbiased only when the elements of $P(\mathbf{S}|D)$ are reasonably well estimated; subsetting (serving as a version of kernel smoothing) has the advantage of increasing the density of information about the cells of this matrix, thus making the estimator approximately unbiased for a much smaller and reasonably sized sample. We find through extensive simulations that this procedure is approximately unbiased, and robust except in very small sample sizes.

## ACKNOWLEDGMENTS

Open source software that implements all the methods described herein is available at http://gking.harvard.edu/va. Our thanks to Bob Black, Ryan Edward, Doug Ewbank, Jianqing Fan, Emmanuela Gakidou, Ken Hill, Kosuke Imai, Henry Kalter, Chris Murray, Stanislava Nikolova, Philip Setel, Kenji Shibuya, Amy Tsui, James Vaupel and Jim Ware for helpful comments and Alan Lopez and Shanon Peter for data assistance. The Tanzania data was provided by the University of Newcastle upon Tyne (UK) as an output of the Adult Morbidity and Mortality Project (AMMP). AMMP was a project of the Tanzanian Ministry of Health, funded by the UK Department for International Development (DFID), and implemented in partnership with the University of Newcastle upon Tyne. Additional funding for the preparation of the data was provided through MEASURE Evaluation, Phase 2, a USAID Cooperative Agreement (GPO-A-00-03-00003-00) implemented by the Carolina Population Center, University of North Carolina at Chapel Hill. This publication was also supported by Grant P10462-109/9903GLOB-2, The Global Burden of Disease 2000 in Aging Populations (P01 AG17625-01), from the United States National Institutes of Health (NIH) National Institute on Aging (NIA) and from the National Science Foundation (SES-0318275, IIS-9874747, DMS-0631652).